\documentclass[%
 aip,
 amsmath,amssymb,
 reprint,%
]{revtex4-1}

\usepackage{graphicx}
\usepackage{dcolumn}
\usepackage{bm}

\usepackage[utf8]{inputenc}
\usepackage[T1]{fontenc}
\usepackage{mathptmx}
\usepackage{etoolbox}
\usepackage{xcolor}

\makeatletter
\def\@email#1#2{%
 \endgroup
 \patchcmd{\titleblock@produce}
  {\frontmatter@RRAPformat}
  {\frontmatter@RRAPformat{\produce@RRAP{*#1\href{mailto:#2}{#2}}}\frontmatter@RRAPformat}
  {}{}
}%
\makeatother
\begin{document}

\preprint{AIP/123-QED}

\title{Trapping a Free-propagating Single-photon into an Atomic Ensemble as a Quantum Stationary Light Pulse}
\author{U-Shin Kim}
\affiliation{Department of Physics, Pohang University of Science and Technology (POSTECH), Pohang 37673, Korea}
\email{usinkimphysics@gmail.com}

\author{Yong Sup Ihn}
\altaffiliation[Current address: ]{Quantum Physics Technology Directorate, Agency for Defense Development, Daejeon 34186, Korea}
\affiliation{Department of Physics, Pohang University of Science and Technology (POSTECH), Pohang 37673, Korea}
 
\author{Chung-Hyun Lee}
\affiliation{Department of Physics, Pohang University of Science and Technology (POSTECH), Pohang 37673, Korea}

\author{Yoon-Ho Kim}
\email{yoonho72@gmail.com}
\thanks{Correspondence to Y.-H. Kim (yoonho72@gmail.com).}
\affiliation{Department of Physics, Pohang University of Science and Technology (POSTECH), Pohang 37673, Korea}

\date{\today}

\begin{abstract}
Efficient photon-photon interaction is one of the key elements for realizing quantum information processing. The interaction, however, must often be mediated through an atomic medium due to the bosonic nature of photons, and the interaction time, which is critically linked to the efficiency, depends on the properties of the atom-photon interaction. While the electromagnetically induced transparency  effect does offer the possibility of photonic quantum memory, it does not enhance the interaction time as it fully maps the photonic state to an atomic state. The stationary light pulse (SLP) effect, on the contrary, traps the photonic state inside an atomic medium with  zero group velocity, opening up the possibility of the enhanced interaction time.  In this work, we report the first experimental demonstration of trapping a free-propagating single-photon into a cold atomic ensemble via the quantum SLP (QSLP) process. We conclusively show that the quantum properties of the single-photon state are preserved well during the QSLP process. Our work paves the way to new approaches for efficient photon-photon interactions, exotic photonic states, and many-body simulations in photonic systems. 
\end{abstract}

\maketitle

\section{Introduction}

\begin{figure*}[t]
\centering
\includegraphics[width=6.69in]{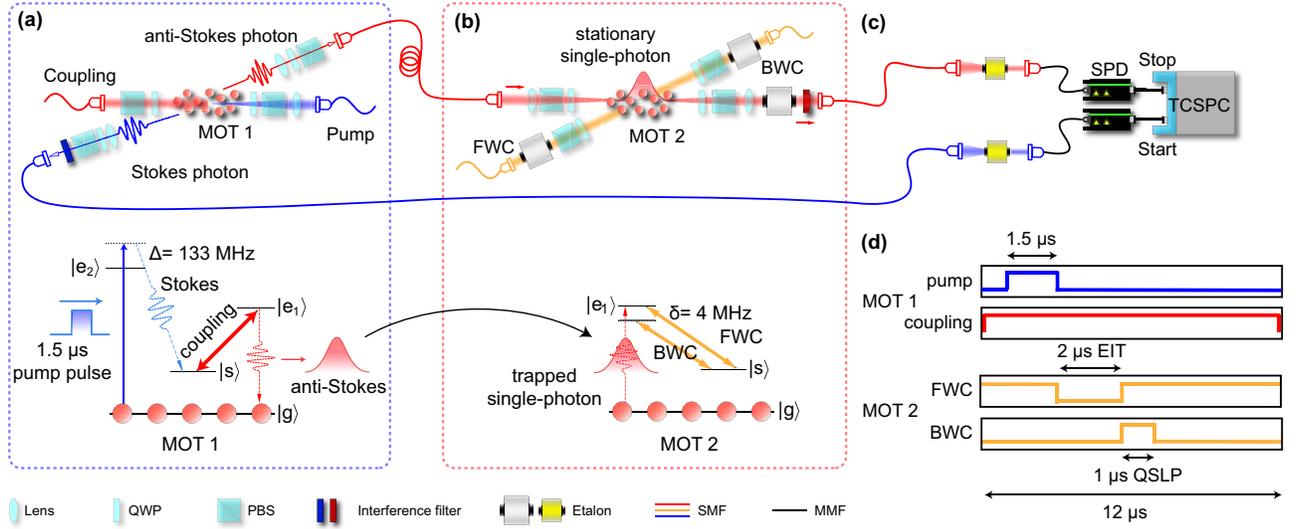}
\caption{Experimental setup. 
(a) Heralded single-photon generation via SFWM in the $^{87}$Rb cold atomic ensemble (MOT 1). The relevant energy levels for  double-$\Lambda$ SFWM are $|g \rangle$=$|5^{2}S_{1/2}, F=1\rangle$, $|s \rangle$ = $|5^{2}S_{1/2}, F=2\rangle$, $|e_{1}\rangle$ = $|5^{2}P_{1/2}, F=2 \rangle$, and $|e_{2}\rangle$ = $|5^{2}P_{3/2}, F=2 \rangle$. 
(b) QSLP trapping of a free-propagating single-photon at MOT 2. The $^{87}$Rb cold atomic ensemble is dressed by the forward coupling (FWC) and backward coupling (BWC) lasers to prepare the atomic medium for the QSLP effect. 
(c) Correlation measurements between the Stokes and the anti-Stokes photons confirm successful trapping of a free-propagating single-photon via the QSLP effect.
(d) The 12 $\mu$s timing sequence of the optical pulses for the experiment. A single experimental run is 100 ms long, out of which 99.7 ms is used to prepare the MOTs and for the duration of 0.3 ms, the 12 $\mu$s long experiment is repeated 25 times. 
}
\label{fig1}
\end{figure*}

Quantum nonlinear optics based on interactions of individual photons is expected to play pivotal roles in many disciplines of science and engineering, notably  in  the emerging field of quantum technology \cite{Dowling03}. Whilst significant advances have been reported in the areas of quantum optics  over the past decades, efficient photon-photon interaction remains to be one of the most challenging problems  and is one of the key elements for realizing quantum information processing  \cite{Chang14}. Since the bosonic nature of photons forces the photon-photon interaction to often be mediated through an atomic medium, the efficiency of the interaction depends critically on the properties of the atom-photon interaction,  and enhancing the atom-photon coupling parameter  for efficient interaction is crucial for various quantum information applications. Recent advances toward this direction have been reported in  cavity quantum electrodynamic systems \cite{Andreas14, Andreas13, Bastian16} and Rydberg atomic systems \cite{Ofer13, Peyronel12, Baur14, Tiarks14}. 

In this regard, the atom-photon interaction time (hence the resulting photon-photon interaction efficiency) may be increased  by arbitrarily slowing down the photons within an atomic medium. While the electromagnetically induced transparency (EIT) effect does offer the mechanisms of slow light and light storage in an atomic medium \cite{Liu01, Phillips01, Cho16}, it  does not significantly enhance the photon-photon interaction time as the photonic state becomes fully mapped to an atomic state when the photon is stored, i.e., zero group velocity  \cite{Fleishhauer02}. The stationary light pulse (SLP) effect, on the contrary, allows trapping of a light pulse inside an atomic medium with  zero group velocity, i.e., the light pulse is not mapped to the atomic state, thus opening up  the possibility of significantly enhancing the photon-photon interaction time  \cite{Wu10, Andre09}. In experiment, the SLP effect occurs in a doubly-driven four-wave mixing scheme in an atomic ensemble, which maintains the electromagnetic field nature under a strong phase-matching condition \cite{Lin09, Everett17}, whereas the EIT memory effect occurs in a singly-driven $\Lambda$ scheme, which does not require any phase-matching condition. 
In former experimental studies on the SLP effect, SLP trapping of a classical light pulse has been reported in atomic systems \cite{Bajcsy03, Lin09, Chen12, Everett17} and generation of a single SLP polariton inside an atomic ensemble has been demonstrated recently \cite{Park18}. However, trapping of a free-propagating single-photon generated in a distant atomic single-photon source,  a crucial step in realizing  photon-photon interaction, has not been reported to date. 

In this work, we report the first experimental demonstration of trapping a free-propagating single-photon into an atomic ensemble via the quantum SLP (QSLP) process. First, a heralded single-photon state is generated in a cold atomic ensemble via the spontaneous four-wave mixing (SFWM) process and sent to another cold atomic ensemble via an optical fiber for QSLP trapping. The formation of the stationary single-photon state, i.e., the demonstration of QSLP trapping, in the second cold atomic ensemble for the free-propagating single-photon from the first cold atomic ensemble is verified by dynamically comparing the QSLP trapping and EIT quantum memory processes. We further demonstrate   that the single-photon nature is  well preserved during the QSLP trapping process in the atomic medium by performing the cross-correlation and the conditional self-correlation measurements. Our work is the first step toward QSLP-based quantum nonlinear optics, including  efficient photon-photon interactions \cite{Iakoupov18}, exotic photonic states \cite{Chang08, Fleischhauer08}, and quantum simulation of relativistic theories \cite{Dimitris13}. 

\section{Experimental setup}

We now elaborate on our experimental setup which is schematically shown in Fig.~\ref{fig1}. First, as shown in Fig.~\ref{fig1}(a), the heralded single-photon state used in the experiment is prepared in a $^{87}$Rb magneto-optical trap (MOT) via the double-$\Lambda$ type SFWM process in which the counter-propagating pump and coupling laser beams generate a pair of energy-time entangled photon pairs, labelled Stokes and anti-Stokes photons \cite{Cho14,Lee16, Ming19,Park17}. The pump laser beam has the pulse width of 1.5 $\mu$s, the power of 70 $\mu$W, and is  133 MHz blue-detuned  from the $|g \rangle$=$|5^{2}S_{1/2}, F=1\rangle$ to $|e_{2}\rangle$ = $|5^{2}P_{3/2}, F=2 \rangle$ transition and the 300 $\mu$W coupling laser is resonant to the $|s \rangle$ = $|5^{2}S_{1/2}, F=2\rangle$ to $|e_{1}\rangle$ = $|5^{2}P_{1/2}, F=2 \rangle$ transition. Both the pump and the coupling laser beams have the polarization state of $\sigma^{-}$. The pump laser beam is focused at the center of MOT 1 to engineer the two-photon correlation function between the Stokes and anti-Stokes photons to be Gaussian-like to improve the efficiency for the subsequent QSLP trapping in MOT 2 \cite{Cho14,Park17,du16}. The detection of the Stokes photon heralds the single-photon state for the anti-Stokes photon and the anti-Stokes photon is sent to MOT 2 via an optical fiber for QSLP trapping.

The schematic for QSLP trapping of a free-propagating single-photon is shown in Fig.~\ref{fig1}(b). The $^{87}$Rb cold atomic ensemble at MOT 2 is dressed by the forward coupling (FWC) and backward coupling (BWC) lasers to prepare the atomic medium for the QSLP effect, where FWC is resonant to the $|g \rangle$ to $|e_{1}\rangle$ transition, and BWC is 4 MHz red-detuned to the transition. BWC is red-detuned by $\delta=2\pi\times4$ MHz from FWC and this detuning, larger than the EIT bandwidth of approximately $2\pi \times 0.6$ MHz, is necessary to suppress higher-order Raman coherence between $\hat{E}^{+}$ ($\hat{E}^{-}$) and BWC (FWC), which hinders formation of QSLP \cite{Lin09,Park18,Moiseev06, Iakoupov16}. QSLP trapping of the anti-Stokes single-photon works as follows: as the anti-Stokes single-photon arrives at MOT 2 along with FWC and BWC laser beams, it is mapped to the dark-state polariton  \cite{Fleischhauer08, Zimmer08},
\begin{equation}
\hat{\psi}=({\hat{E}}^{+}\mathrm{cos}\phi+{\hat{E}}^{-}\mathrm{sin}\phi)\,\mathrm{cos}\theta-\hat{{S}}\,\mathrm{sin}\theta,
\end{equation}
where the first two terms are the photonic components and the third term is the atomic component. Here, ${\hat{E}^{+}}$ and ${\hat{E}^{-}}$ represent the forward and backward propagating electromagnetic field operators of the single-photon, respectively,  ${\hat{S}}$ represents the collective atomic state defined as $\hat{{S}}=\frac{1}{\sqrt{N}}\sum_{i=1}^{N}|g\rangle_{ii}\langle s|$ with $|g\rangle_i$ and $|s\rangle_i$ representing the hyperfine states in Fig.~\ref{fig1}(b) for the $i$-th atom, and $N$ is the number of atoms \cite{Fleischhauer00}. The photonic and atomic components of the dark state polariton in Eq.~(1) are characterized by the two mixing angles $\theta$ and  $\phi$ defined as  $\mathrm{tan}^{2}\theta=g^2N/\Omega^2$ and $\mathrm{tan}^{2}\phi=\Omega_{\mathrm{BWC}}/\Omega_{\mathrm{FWC}}$, where $g$ is the atom-photon coupling constant, $\Omega_{\mathrm{FWC}}$ and $\Omega_{\mathrm{BWC}}$ are the Rabi frequencies of the respective FWC and BWC beams, and $\Omega^{2}=\Omega_{\mathrm{FWC}}^{2}+\Omega_{\mathrm{BWC}}^{2}$. The group velocity of the dark state polariton is given as  \cite{Hansen07},
\begin{equation}
v_g=c_{0}\, \mathrm{cos}^{2}\theta \, \mathrm{cos}2\phi,
\end{equation}
where $c_{0}$ is the speed of light in vacuum. It is interesting to note that at $\phi=45^\circ$, i.e., $\Omega_{\mathrm{FWC}}  = \Omega_{\mathrm{BWC}}$, the group velocity of the dark state polariton becomes zero even though the photonic components of the dark state polariton are non-vanishing, leading to trapping of the single-photon within the atomic medium via the QSLP effect. 

The measurement setup, which consists of  the single-photon detectors (SPD) and a time-correlated single-photon counting (TCSPC) device, is shown in Fig.~\ref{fig1}(c).
The temporal correlation measurements  between the Stokes and the anti-Stokes photons are used to confirm successful trapping of a free-propagating single-photon via the QSLP effect. The timing sequence of the optical pulses for the experiment is shown Fig.~\ref{fig1}(d). A single experimental run is 100 ms long, out of which 99.7 ms is used to prepare the MOTs (not shown) and for the duration of 0.3 ms, the 12 $\mu$s long experiment shown in Fig.~\ref{fig1}(d) is repeated 25 times.

To verify  formation of the stationary single-photon state in the atomic medium, it is necessary to clearly distinguish the QSLP process from the EIT memory process at MOT 2. Thus, a 2 $\mu$s long EIT memory phase is first introduced just before the 1 $\mu$s long QSLP trapping phase, see Fig.~\ref{fig1}(d). As the anti-Stokes single-photon wave packet arrives at MOT 2 with FWC turned on, it gets slowed down in the atomic medium and once the single-photon wave packet has fully entered the atomic medium, FWC is turned off for the duration of 2 $\mu$s to fully map the single-photon into the atomic state ${\hat{S}}$. After the EIT memory phase of 2 $\mu$s, FWC is turned back on to reversely map the atomic state to the photonic state, retrieving the free-propagating single-photon from the EIT memory. For QSLP trapping of the retrieved single-photon from the EIT memory phase,  both FWC and BWC are simultaneously turned on immediately after the release of the single-photon. After the QSLP trapping phase of 1 $\mu$s, BWC is turned off to release the trapped single-photon in the same direction as that of the original anti-Stokes single photon from MOT 1.

For QSLP trapping of the  anti-Stokes single-photon at MOT 2, the following phase-matching condition needs to be satisfied  \cite{Lin09, Everett17},
\begin{equation}
\vec{k}^{+}_{as}-\vec{k}_{FWC}=\vec{k}^{-}_{as}-\vec{k}_{BWC},
\end{equation}
where $\vec{k}_{FWC}$ ($\vec{k}_{BWC}$) represents the wave vector of  FWC (BWC) and $\vec{k}^{+}_{as}$ ($\vec{k}^{-}_{as}$) represents the wave vector of the forward (backward) propagating component of the anti-Stokes photon in Eq.~(1). Considering the involved energy levels shown in Fig.~\ref{fig1},  Eq.~(3) requires the angle between the input anti-Stokes single-photon and FWC/BWC beams to be 0.345$^{\circ}$ to minimize the phase-mismatching and thus, the leakage of the QSLP process while trapping single-photons. Both FWC and BWC are $\sigma^{-}$ polarized and are 300 $\mu$W in power.

\section{Results}

\begin{figure}[t]
\centering
\includegraphics[width=3.2in]{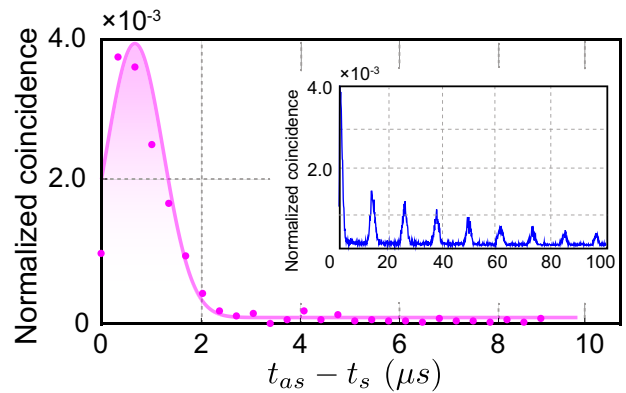}
\caption{The waveform of the heralded single-photon state for the anti-Stokes photon. The solid line is a Gaussian fit with FWHM of 0.57 $\mu$s. The inset shows the extended TCSPC histogram up to 100 $\mu$s. The normalized cross-correlation is $g^{(2)}(0)$=11.12$\pm$0.07 from the raw data and $g^{(2)}(0)$=11.67$\pm$0.07 from the noise-subtracted data. The error bar is not shown in the plot since the size of the error bar is smaller than the size of data points.}
\label{fig2}
\end{figure}

The waveform of the  single-photon state for the anti-Stokes photon, heralded by the detection of the Stokes photon, generated in MOT 1 is shown in Fig.~\ref{fig2}. The TCSPC histogram  is normalized by the single count of the Stokes photon to show the number of the heralded anti-Stokes single-photon per  heralding event. The waveform of the heralded single-photon state is engineered to have a Gaussian-like shape with the full-width half maximum (FWHM) of 0.57 $\mu$s by focusing the pump beam at the center of MOT 1 \cite{du16,Wang19}. The inset in Fig.~\ref{fig2} shows the extended TCSPC histogram up to 100 $\mu$s. 
The first peak is due to the cross-correlation between the Stokes and the anti-Stokes photons generated in the same pulse sequence shown in Fig.~\ref{fig1}(d)  and the following smaller peaks are due to the cross-correlation between photons from subsequent pulse sequences. The decay of the smaller peaks with the increasing delay time $t_{as} - t_s$ is due to low coupling beam power, which is insufficient to excite the atom   from $|s\rangle$ to $|e_{1}\rangle$ with  probability unity. An atom which had not been excited in a pulse sequence may still be excited in the following sequences, explaining the decay. 
The normalized cross-correlation $g^{(2)}(0)$ is obtained by dividing the total count of the first peak summed from 0 to 3.5 $\mu$s  by that of the last peak in the inset with the same temporal window (see Supplementary material). The normalized cross-correlation is calculated to be $g^{(2)}(0)$=11.12$\pm$0.07 from the raw data and $g^{(2)}(0)$=11.67$\pm$0.07 from the noise-subtracted data. The primary noise source in our experiment is the FWC leakage which contributes to the DC-offset in the TCSPC histogram. The noise-subtracted data is obtained by subtracting the DC-offset, which is evaluated by averaging the plateau regions between the subsequent peaks in the inset of Fig.~\ref{fig2},  from the relevant TCSPC peaks.

\begin{figure}[t]
\centering
\includegraphics[width=3.2in]{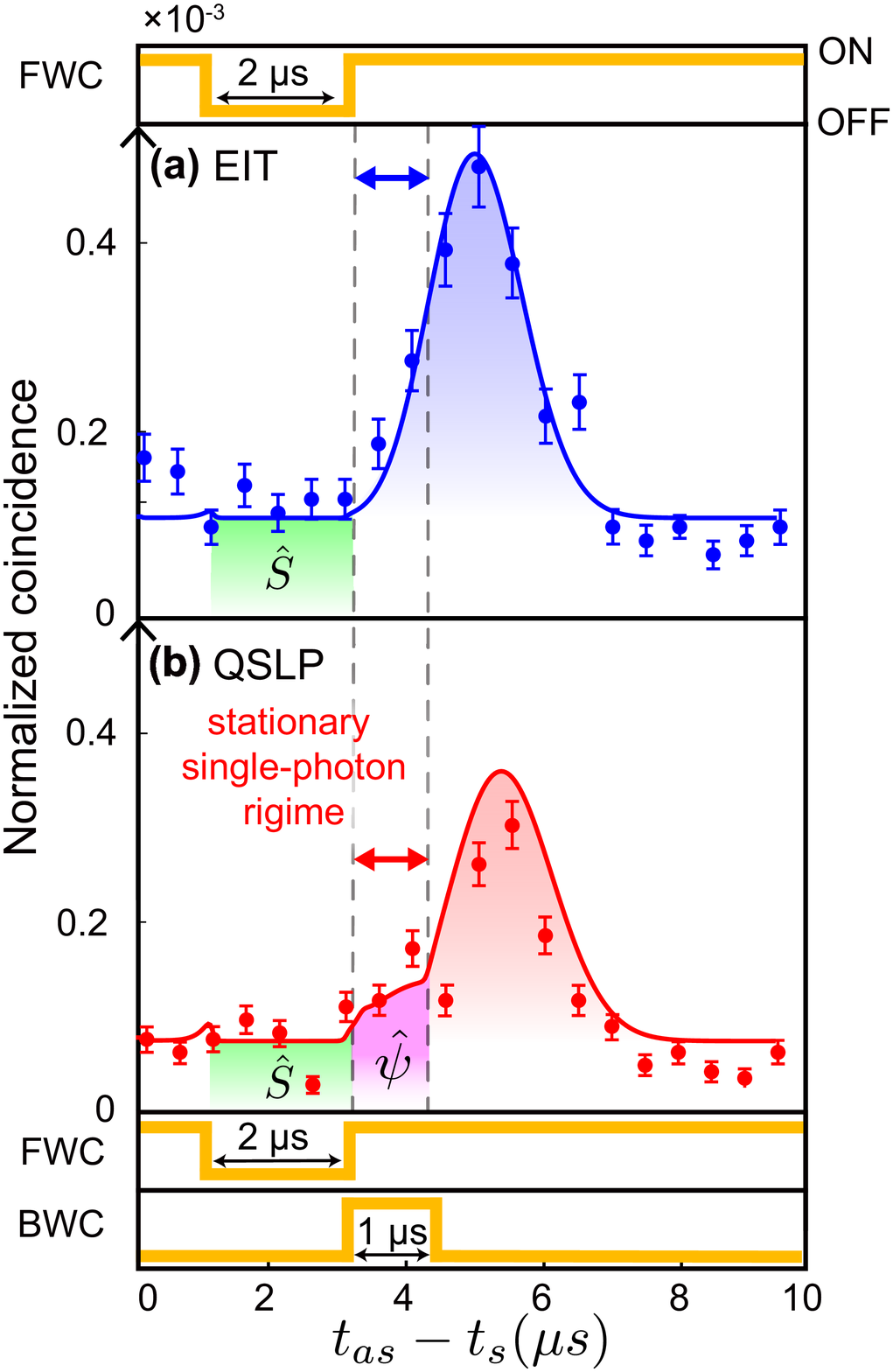}
\caption{The normalized  TCSPC histograms for (a) the EIT memory  and (b) the QSLP trapping. (a) The single-photon wave packet released from  2 $\mu$s EIT memory.
(b) The single-photon wave packet released from 2 $\mu$s EIT memory followed by 1 $\mu$s QSLP trapping. 
The solid lines are due to Eq.~(C1) $\sim$ Eq.~(C7) with the following parameters: the optical depth (OD) $=100$, the ground state decoherence $\gamma_{gs}$= $2\pi\times60$ kHz, and the Rabi frequencies of  FWC and BWC are $2\pi\times6.0$ MHz and $2\pi\times4.2$ MHz, respectively.}
\label{fig3}
\end{figure}

The experimental data for EIT memory and QSLP trapping are shown in Fig.~\ref{fig3}. First, the TCSPC histogram for the heralded single-photon wave packet released from  2 $\mu$s EIT memory is shown in Fig.~\ref{fig3}(a). During the 2 $\mu$s EIT memory duration, the heralded single-photon wave packet is fully mapped to the collective atomic state (green area). Then, as FWC is turned on, the collective atomic state is retrieved back to the free-propagating single photon (blue area). The retrieval efficiency of the 2 $\mu$s EIT memory is measured to be 19.1 \% from accumulated counts of the noise-subtracted traces. Note that the energy loss mainly comes from the narrow EIT bandwidth of MOT 2, which is set for a proper group delay of the EIT memory process. Next,  the single-photon wave packet released from 2 $\mu$s EIT memory followed by 1 $\mu$s QSLP trapping is shown in Fig.~\ref{fig3}(b). In contrast to Fig.~\ref{fig3}(a), when both FWC and BWC are simultaneously turned on for an additional 1 $\mu$s, photon emission from MOT 2 is suppressed  (the area marked with $\hat \psi$) and the peak arrival time delayed by that duration compared to the EIT memory case in Fig.~\ref{fig3}(a).  The suppressed and delayed emission of the single-photon wave packet in Fig.~\ref{fig3}(b) clearly verify QSLP trapping of the heralded anti-Stokes single-photon wave packet. The release efficiency of the 1 $\mu$s QSLP trapping process compared to the 2 $\mu$s EIT memory process is measured to be 47.8 \% from accumulated counts of the noise-subtracted traces, which corresponds to the exponential energy decay time of 1.35 $\mu$s.
 
 \begin{figure}[t]
\centering
\includegraphics[width=3.4in]{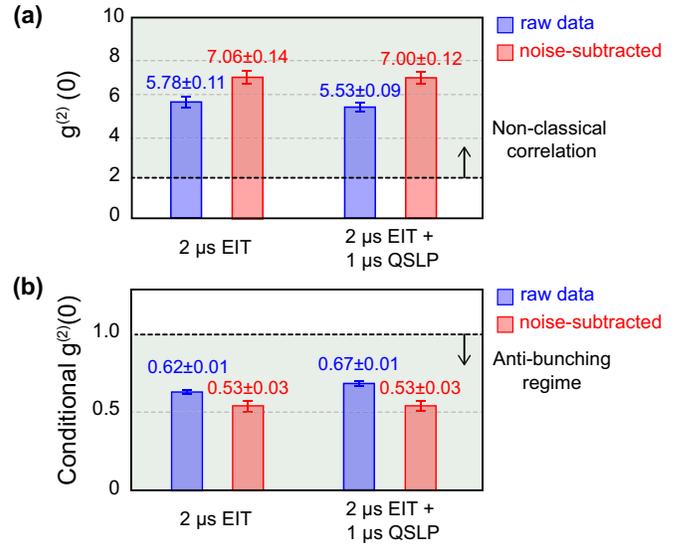}
\caption{(a) Normalized cross-correlation $g^{(2)}(0)$ between the Stokes and the anti-Stokes photons after 2 $\mu$s EIT memory and 2 $\mu$s EIT memory followed by 1 $\mu$s QSLP trapping. (b) Normalized  self-correlation of the heralded anti-Stokes photon after 2 $\mu$s EIT memory and 2 $\mu$s EIT memory followed by 1 $\mu$s QSLP trapping. The conditional  $g^{(2)}(0)$ value of the heralded anti-Stokes photon before EIT memory is obatined to be  0.35 $\pm$ 0.01 from  the raw data and 0.33 $\pm$ 0.01 from the noise-subtracted data.}
\label{fig4}
\end{figure}

To verify that the quantum nature of the single-photon state is preserved during the QSLP trapping process, we measured the cross-correlation between the Stokes and the anti-Stokes photons as well as the self-correlation of the heralded anti-Stokes photon after the EIT memory and QSLP trapping processes. For these measurements, coincidence counts are accumulated for the time windows between 3.3 $\mu$s and 7.0 $\mu$s in Fig.~\ref{fig3}(a) and 4.3 $\mu$s and 8.0 $\mu$s in Fig.~\ref{fig3}(b) and the experimental data are shown in Fig.~\ref{fig4} (see Appendix B). The normalized cross-correlation $g^{(2)}(0)$ between the Stokes and the anti-Stokes photons after 2 $\mu$s EIT memory and 2 $\mu$s EIT memory followed by 1 $\mu$s QSLP trapping are shown in Fig.~\ref{fig4}(a) and the data clearly confirm that the non-classical nature of the correlation is well-preserved during the QSLP trapping process. Moreover, the normalized self-correlation of the heralded anti-Stokes photon after 2 $\mu$s EIT memory and 2 $\mu$s EIT memory followed by 1 $\mu$s QSLP trapping shown in Fig.~\ref{fig4}(b) confirms preservation of the quantum nature of the heralded anti-Stokes photon. Note that the loss of quantum correlations after the EIT memory and QSLP process compared to the SFWM photon pair is due to the narrow EIT window of MOT 2, which produces a frequency-dependent loss to the anti-Stokes photon by its Lorentizian shape. The frequency-dependent loss degrades the energy-time entanglement between Stokes and anti-Stokes photon since the entanglement lies at the frequency modes of two photons. Hence, the loss of quantum correlation can be reduced by implementing a single-photon source with a narrower bandwidth than the EIT window of MOT 2.

\section{conclusion}

In conclusion, we have reported the first experimental demonstration of trapping a free-propagating single-photon into a cold atomic ensemble via the quantum stationary light pulse process. Successful QSLP trapping of a single-photon has been verified by experimentally comparing the two processes: EIT memory  and  EIT memory followed by QSLP trapping. The suppressed emission during the duration of QSLP and the delayed emission of the single-photon wave packet observed via the time-correlation measurements clearly show  QSLP trapping of the single-photon wave packet in the atomic ensemble. We have also shown that the quantum nature of the single-photon state is preserved during the QSLP trapping. This result indicates that the QSLP trapping process may be used for efficient photon-photon interaction such as photon blockade without an optical cavity, where a single-photon trapped in an atomic ensemble as a QSLP drastically changes optical properties of the atomic ensemble \cite{Iakoupov16}. Our work is the first step toward QSLP-based quantum nonlinear optics, including  efficient photon-photon interactions \cite{Iakoupov18}, exotic photonic states \cite{Chang08, Fleischhauer08}, and quantum simulation of relativistic theories \cite{Dimitris13}. 

\section*{Supplementary material}
See supplementary material for the detailed evaluation of quantum correlations from experimental TSCPS histograms.

\begin{acknowledgments}
This work was funded in part by the National Research Foundation of Korea (Grant No. 2019R1A2C3004812) and ITRC support program (IITP-2021-2020-0-01606).
\end{acknowledgments}

\section*{author declarations}

\subsection*{Conflict of interest}
The authors have no conflicts to disclose.

\subsection*{Data Availability}
The data that support the findings of this study are available from the corresponding author upon reasonable request. 

\appendix

\section{Noise filtering setup}

Due to the phase matching condition for QSLP trapping between the input anti-Stokes photon and FWC/BWC lasers in Eq.~(3),  the anti-Stokes photon emitted from MOT 2 has the angle of $0.345^\circ$ with respect to FWC/BWC lasers. Therefore, filtering out the FWC leakage completely is important in observing QSLP trapping of the anti-Stokes photon. 
The anti-Stokes photon emitted from MOT 2 is first  filtered by a planar-planar etalon of 1 GHz bandwidth before the single-mode fiber and then by a planar-concave etalon of 215 MHz bandwidth after the fiber. The free-space planar-planar etalon helps to filter out FWC from the emitted anti-Stokes photon spatially due to their different collimating conditions: the anti-Stokes photon is focused to have a diameter of 250 $\mu$m at the atomic ensemble (MOT 2) while FWC is collimated with a diameter of 1.8 mm. To further filter out the FWC leakage, planar-planar etalons of 80 MHz and 1 GHz bandwidths are introduced to FWC and BWC, respectively. These planar-planar etalons help to filter out the long frequency tails of the lasers used in our experiment.

\section{Calculation of conditional $g^{(2)}(0)$}

The quantum state of the Stokes and the anti-Stokes photons from the SFWM process can be described as a two-mode squeezed state in the Fock state basis as \cite{Hongyi02, Cochrane02}, 
\begin{equation}
|\psi\rangle=\sqrt{1-\eta}\sum_{n=0}^{\infty}\eta^{n/2}|n\rangle_{s}|n\rangle_{as},
\label{state}
\end{equation}
where the subscripts $s$ and $as$ represent the Stokes and anti-Stokes photons, respectively. The normalized cross-correlation between the Stokes and the anti-Stokes photons is written as
\begin{equation}
g^{(2)}_{s, as}(0)=\frac{\langle \hat{a}_{s}^{\dagger} \hat{a}_{as}^{\dagger} \hat{a}_{as} \hat{a}_{s} \rangle}{\langle \hat{a}_{as}^{\dagger} \hat{a}_{as}\rangle \langle \hat{a}_{s}^{\dagger} \hat{a}_{s}\rangle},\nonumber
\end{equation}
and, for the quantum state in Eq.~(\ref{state}), $\langle \hat{a}_{as}^{\dagger} \hat{a}_{as}\rangle =\langle \hat{a}_{s}^{\dagger} \hat{a}_{s}\rangle 
= (1-\eta)\sum_{n=0}^{\infty}\eta^{n}n = \eta/ (1-\eta)^{2}$ and 
$ \langle \hat{a}_{s}^{\dagger} \hat{a}_{as}^{\dagger} \hat{a}_{as} \hat{a}_{s} \rangle =(1-\eta)\sum_{n=0}^{\infty}\eta^{n}n^{2} =  \eta(\eta+1) / (1-\eta)^{2}$.
The cross-correlation $g^{(2)}_{s, as}(0)$, therefore, is found to be, 
\begin{equation}
g^{(2)}_{s, as}(0)=1+\frac{1}{\eta}.
\label{g_cross}
\end{equation}

The conditional self-correlation $g_{c}^{(2)}(0)$ for the anti-Stokes photon heralded by the detection of the Stokes photon is given by,
\begin{equation}
g_{c}^{(2)}(0)=\frac{\langle \hat{a}_{s}^{\dagger} \hat{a}_{as1}^{\dagger} \hat{a}_{as2}^{\dagger} \hat{a}_{as2} \hat{a}_{as1} \hat{a}_{s}\rangle}{\langle \hat{a}_{s}^{\dagger} \hat{a}_{as1}^{\dagger} \hat{a}_{as1} \hat{a}_{s} \rangle \langle \hat{a}_{s}^{\dagger} \hat{a}_{as2}^{\dagger} \hat{a}_{as2} \hat{a}_{s} \rangle},
\end{equation}
where $\hat{a}_{as1}$ and $\hat{a}_{as2}$ refer to the annihilation operators for the output modes a symmetric beam splitter (BS) on which the anti-Stokes photon is incident. For the input modes of the BS, $\hat{a}_{as}$ and $\hat{a}_{vac}$, respectively, refer to the annihilation operators for for the mode that the anti-Stokes photon is sent to and that of the vacuum. Note that the following relation holds:  $\hat{a}_{as1}=(\hat{a}_{as}-\hat{a}_{vac})/\sqrt{2}$ and $\hat{a}_{as2}=(\hat{a}_{as}+\hat{a}_{vac})/\sqrt{2}$. For the quantum state in Eq.~(\ref{state}), $\langle \hat{a}_{s}^{\dagger} \hat{a}_{as1}^{\dagger} \hat{a}_{as1} \hat{a}_{s} \rangle =\frac{1}{2}(1-\eta)\sum_{n=0}^{\infty}\eta^{n}n^{2}$ and $\langle \hat{a}_{s}^{\dagger} \hat{a}_{as1}^{\dagger} \hat{a}_{as2}^{\dagger} \hat{a}_{as2} \hat{a}_{as1} \hat{a}_{s}\rangle = \frac{1}{4}(1-\eta)\sum_{n=0}^{\infty}\eta^{n}n^{2}(n-1)$, which lead to,
\begin{equation}
g_{c}^{(2)}(0)=\frac{2\eta(\eta+2)}{(\eta+1)^{2}}.
\end{equation}
The normalized conditional self-correlation $g_{c}^{(2)}(0)$ for the heralded anti-Stokes photon, thus, can be obtained from the relation in Eq.~(\ref{g_cross}),  $\eta=1/(g^{(2)}_{s, as}(0)-1)$.

\section{Numerical simulation of QSLP}
We analyzed the experimental results by numerically solving the partial differential equations for the stationary light pulse in a cold atomic medium \cite{Chen12},
\begin{gather}
\frac{\partial \rho_{eg}^{+}}{\partial t} = \frac{i}{2} E_{as}^{+} + \frac{i}{2}( \rho_{sg}^{0} \Omega_{FWC}+  \rho_{sg}^{-+} \Omega_{BWC})-\frac{\Gamma}{2} \rho_{eg}^{+}, \\
\frac{\partial \rho_{eg}^{-}}{\partial t} = \frac{i}{2} E_{as}^{-} + \frac{i}{2}(  \rho_{sg}^{0} \Omega_{BWC}+  \rho_{sg}^{+-} \Omega_{FWC})-\left (\frac{\Gamma}{2} - i \delta\right )\rho_{eg}^{-}, \\
\frac{\partial \rho_{sg}^{+-}}{\partial t} = \frac{i}{2} \Omega_{FWC}^{*} \rho_{eg}^{-}-\left (\gamma_{gs}'-i\delta \right )\rho_{sg}^{+-}, \\
\frac{\partial \rho_{sg}^{-+}}{\partial t} = \frac{i}{2} \Omega_{BWC}^{*} \rho_{eg}^{+}-\left (\gamma_{gs}' +i\delta \right )\rho_{sg}^{-+}, \\
\frac{\partial \rho_{sg}^{0}}{\partial t} = \frac{i}{2}(\Omega_{FWC}^{*}   \rho_{eg}^{+}+\Omega_{BWC}^{*}  \rho_{eg}^{-})-\gamma_{gs} \rho_{sg}^{0}, \\
\left ( \frac{1}{c_{0}}\frac{\partial}{\partial t}+\frac{\partial}{\partial z}\right )E_{as}^{+} = i\frac{\mathrm{OD} \Gamma}{2L}\rho^{+}_{eg},\\
\left ( \frac{1}{c_{0}}\frac{\partial}{\partial t}-\frac{\partial}{\partial z}+i\Delta k\right )E_{as}^{-} = i\frac{\mathrm{OD} \Gamma}{2L}\rho^{-}_{eg},
\end{gather}
where $\rho_{ij}$ is the $ij$-th element of the atomic density operator, i.e., the slowly-varying amplitude of the coherence between states $\left | i \right >$ and $\left | j \right >$, 
$\Gamma$ is the radiative decay rate of level $\left | e\right >$, 
$\gamma_{gs}$ and $\gamma'_{gs}$ respectively are the ground state dephasing rates for  $\rho_{gs}^{0}$ and  $\rho_{gs}^{+-}$ (as well as $\rho_{gs}^{-+}$ ), 
$\delta$ is the experimentally-chosen detuning in Fig. \ref{fig1} for suppressing the higher-order Raman coherence, 
$c_{0}$ is the light speed in vacuum, 
$\mathrm{OD}$ is the optical depth of the atomic medium of MOT 2,    
$L$ is the length of the medium, and 
 $\Delta k = \left ( \vec{k}_{as}-\vec{k}^{+}_{FWC}-\vec{k}^{-}_{BWC}\right ) \times \hat{z}-k = k(\cos 2\Theta -1)$,  where $\Theta$ is the angle between the FWC/BWC beams and anti-Stokes photon  \cite{Danielle04}. The FWC/BWC lasers are time-depenent so that $\Omega_{FWC} = \Omega_{FWC}^{0} \times Q^{+}(t)$ and $\Omega_{BWC} = \Omega_{BWC}^{0} \times Q^{-}(t)$, where $Q^{+}(t)$ and $Q^{-}(t)$ are the experimental timing sequences  in Fig.~\ref{fig1}(d). For the anti-Stokes photon, the Gaussian shape shown in Fig. \ref{fig2} is used as the initial condition for numerically solving Eq.~(C1) to Eq.~(C7) using the finite difference method 
 to reveal the dynamics of QSLP \cite{Dehghan99,Curtiss52}.  The experimental data agree well with the numerical simulation results, as shown in Fig.~\ref{fig3}, using the following parameters: $\Omega_{FWC}^{0} = 2\pi\times6.0$ MHz, $\Omega_{BWC}^{0} = 2\pi \times4.2$ MHz, $\mathrm{OD}$ = 100, $\gamma_{gs} = \gamma'_{gs} = 2\pi\times60$ kHz, $L$ = 10 mm, and $\Delta k = 5$ m$^{-1}$.

\section*{references}
\nocite{*}
\bibliography{aipsamp}

\providecommand{\noopsort}[1]{}\providecommand{\singleletter}[1]{#1}%
\begin{thebibliography}{40}%
\makeatletter
\providecommand \@ifxundefined [1]{%
 \@ifx{#1\undefined}
}%
\providecommand \@ifnum [1]{%
 \ifnum #1\expandafter \@firstoftwo
 \else \expandafter \@secondoftwo
 \fi
}%
\providecommand \@ifx [1]{%
 \ifx #1\expandafter \@firstoftwo
 \else \expandafter \@secondoftwo
 \fi
}%
\providecommand \natexlab [1]{#1}%
\providecommand \enquote  [1]{``#1''}%
\providecommand \bibnamefont  [1]{#1}%
\providecommand \bibfnamefont [1]{#1}%
\providecommand \citenamefont [1]{#1}%
\providecommand \href@noop [0]{\@secondoftwo}%
\providecommand \href [0]{\begingroup \@sanitize@url \@href}%
\providecommand \@href[1]{\@@startlink{#1}\@@href}%
\providecommand \@@href[1]{\endgroup#1\@@endlink}%
\providecommand \@sanitize@url [0]{\catcode `\\12\catcode `\$12\catcode
  `\&12\catcode `\#12\catcode `\^12\catcode `\_12\catcode `\%12\relax}%
\providecommand \@@startlink[1]{}%
\providecommand \@@endlink[0]{}%
\providecommand \url  [0]{\begingroup\@sanitize@url \@url }%
\providecommand \@url [1]{\endgroup\@href {#1}{\urlprefix }}%
\providecommand \urlprefix  [0]{URL }%
\providecommand \Eprint [0]{\href }%
\providecommand \doibase [0]{http://dx.doi.org/}%
\providecommand \selectlanguage [0]{\@gobble}%
\providecommand \bibinfo  [0]{\@secondoftwo}%
\providecommand \bibfield  [0]{\@secondoftwo}%
\providecommand \translation [1]{[#1]}%
\providecommand \BibitemOpen [0]{}%
\providecommand \bibitemStop [0]{}%
\providecommand \bibitemNoStop [0]{.\EOS\space}%
\providecommand \EOS [0]{\spacefactor3000\relax}%
\providecommand \BibitemShut  [1]{\csname bibitem#1\endcsname}%
\let\auto@bib@innerbib\@empty
\bibitem [{\citenamefont {Dowling}\ and\ \citenamefont
  {Milburn}(2003)}]{Dowling03}%
  \BibitemOpen
  \bibfield  {author} {\bibinfo {author} {\bibfnamefont {J.~P.}\ \bibnamefont
  {Dowling}}\ and\ \bibinfo {author} {\bibfnamefont {G.~J.}\ \bibnamefont
  {Milburn}},\ }\bibfield  {title} {\enquote {\bibinfo {title} {Quantum
  technology: the second quantum revolution},}\ }\href@noop {} {\bibfield
  {journal} {\bibinfo  {journal} {Phil. Trans. R. Soc. Lond. A}\ }\textbf
  {\bibinfo {volume} {361}},\ \bibinfo {pages} {1655--1674} (\bibinfo {year}
  {2003})}\BibitemShut {NoStop}%
\bibitem [{\citenamefont {Chang}, \citenamefont {Vuleti\'{c}},\ and\
  \citenamefont {Lukin}(2014)}]{Chang14}%
  \BibitemOpen
  \bibfield  {author} {\bibinfo {author} {\bibfnamefont {D.~E.}\ \bibnamefont
  {Chang}}, \bibinfo {author} {\bibfnamefont {V.}~\bibnamefont {Vuleti\'{c}}},
  \ and\ \bibinfo {author} {\bibfnamefont {M.~D.}\ \bibnamefont {Lukin}},\
  }\bibfield  {title} {\enquote {\bibinfo {title} {Quantum nonlinear optics -
  photon by photon},}\ }\href@noop {} {\bibfield  {journal} {\bibinfo
  {journal} {Nature Photonics}\ }\textbf {\bibinfo {volume} {8}},\ \bibinfo
  {pages} {685--694} (\bibinfo {year} {2014})}\BibitemShut {NoStop}%
\bibitem [{\citenamefont {Reiserer}\ \emph {et~al.}(2014)\citenamefont
  {Reiserer}, \citenamefont {Kalb}, \citenamefont {Rempe},\ and\ \citenamefont
  {Ritter}}]{Andreas14}%
  \BibitemOpen
  \bibfield  {author} {\bibinfo {author} {\bibfnamefont {A.}~\bibnamefont
  {Reiserer}}, \bibinfo {author} {\bibfnamefont {N.}~\bibnamefont {Kalb}},
  \bibinfo {author} {\bibfnamefont {G.}~\bibnamefont {Rempe}}, \ and\ \bibinfo
  {author} {\bibfnamefont {S.}~\bibnamefont {Ritter}},\ }\bibfield  {title}
  {\enquote {\bibinfo {title} {A quantum gate between a flying optical photon
  and a single trapped atom},}\ }\href@noop {} {\bibfield  {journal} {\bibinfo
  {journal} {Nature}\ }\textbf {\bibinfo {volume} {508}},\ \bibinfo {pages}
  {237--240} (\bibinfo {year} {2014})}\BibitemShut {NoStop}%
\bibitem [{\citenamefont {Reiserer}, \citenamefont {Ritter},\ and\
  \citenamefont {Rempe}(2013)}]{Andreas13}%
  \BibitemOpen
  \bibfield  {author} {\bibinfo {author} {\bibfnamefont {A.}~\bibnamefont
  {Reiserer}}, \bibinfo {author} {\bibfnamefont {S.}~\bibnamefont {Ritter}}, \
  and\ \bibinfo {author} {\bibfnamefont {G.}~\bibnamefont {Rempe}},\ }\bibfield
   {title} {\enquote {\bibinfo {title} {Nondestructive detection of an optical
  photon},}\ }\href@noop {} {\bibfield  {journal} {\bibinfo  {journal}
  {Science}\ }\textbf {\bibinfo {volume} {342}},\ \bibinfo {pages} {1349--1351}
  (\bibinfo {year} {2013})}\BibitemShut {NoStop}%
\bibitem [{\citenamefont {Hacker}\ \emph {et~al.}(2016)\citenamefont {Hacker},
  \citenamefont {Welte}, \citenamefont {Rempe},\ and\ \citenamefont
  {Ritter}}]{Bastian16}%
  \BibitemOpen
  \bibfield  {author} {\bibinfo {author} {\bibfnamefont {B.}~\bibnamefont
  {Hacker}}, \bibinfo {author} {\bibfnamefont {S.}~\bibnamefont {Welte}},
  \bibinfo {author} {\bibfnamefont {G.}~\bibnamefont {Rempe}}, \ and\ \bibinfo
  {author} {\bibfnamefont {S.}~\bibnamefont {Ritter}},\ }\bibfield  {title}
  {\enquote {\bibinfo {title} {A photon-photon quantum gate based on a single
  atom in an optical resonator},}\ }\href@noop {} {\bibfield  {journal}
  {\bibinfo  {journal} {Nature}\ }\textbf {\bibinfo {volume} {536}},\ \bibinfo
  {pages} {193--196} (\bibinfo {year} {2016})}\BibitemShut {NoStop}%
\bibitem [{\citenamefont {Firstenberg}\ \emph {et~al.}(2013)\citenamefont
  {Firstenberg}, \citenamefont {Peyronel}, \citenamefont {Liang}, \citenamefont
  {Gorshkov}, \citenamefont {Lukin},\ and\ \citenamefont
  {Vuleti\'{c}}}]{Ofer13}%
  \BibitemOpen
  \bibfield  {author} {\bibinfo {author} {\bibfnamefont {O.}~\bibnamefont
  {Firstenberg}}, \bibinfo {author} {\bibfnamefont {T.}~\bibnamefont
  {Peyronel}}, \bibinfo {author} {\bibfnamefont {Q.~Y.}\ \bibnamefont {Liang}},
  \bibinfo {author} {\bibfnamefont {A.~V.}\ \bibnamefont {Gorshkov}}, \bibinfo
  {author} {\bibfnamefont {M.~D.}\ \bibnamefont {Lukin}}, \ and\ \bibinfo
  {author} {\bibfnamefont {V.}~\bibnamefont {Vuleti\'{c}}},\ }\bibfield
  {title} {\enquote {\bibinfo {title} {Attractive photons in a quantum
  nonlinear medium},}\ }\href@noop {} {\bibfield  {journal} {\bibinfo
  {journal} {Nature}\ }\textbf {\bibinfo {volume} {502}},\ \bibinfo {pages}
  {71--75} (\bibinfo {year} {2013})}\BibitemShut {NoStop}%
\bibitem [{\citenamefont {Peyronel}\ \emph {et~al.}(2012)\citenamefont
  {Peyronel}, \citenamefont {Firstenberg}, \citenamefont {Liang}, \citenamefont
  {Hofferberth}, \citenamefont {Gorshkov}, \citenamefont {Pohl}, \citenamefont
  {Lukin},\ and\ \citenamefont {Vuleti\'{c}}}]{Peyronel12}%
  \BibitemOpen
  \bibfield  {author} {\bibinfo {author} {\bibfnamefont {T.}~\bibnamefont
  {Peyronel}}, \bibinfo {author} {\bibfnamefont {O.}~\bibnamefont
  {Firstenberg}}, \bibinfo {author} {\bibfnamefont {Q.~Y.}\ \bibnamefont
  {Liang}}, \bibinfo {author} {\bibfnamefont {S.}~\bibnamefont {Hofferberth}},
  \bibinfo {author} {\bibfnamefont {A.~V.}\ \bibnamefont {Gorshkov}}, \bibinfo
  {author} {\bibfnamefont {T.}~\bibnamefont {Pohl}}, \bibinfo {author}
  {\bibfnamefont {M.~D.}\ \bibnamefont {Lukin}}, \ and\ \bibinfo {author}
  {\bibfnamefont {V.}~\bibnamefont {Vuleti\'{c}}},\ }\bibfield  {title}
  {\enquote {\bibinfo {title} {Quantum nonlinear optics with single photons
  enabled by strongly interacting atoms},}\ }\href@noop {} {\bibfield
  {journal} {\bibinfo  {journal} {Nature}\ }\textbf {\bibinfo {volume} {488}},\
  \bibinfo {pages} {57--60} (\bibinfo {year} {2012})}\BibitemShut {NoStop}%
\bibitem [{\citenamefont {Baur}\ \emph {et~al.}(2014)\citenamefont {Baur},
  \citenamefont {Tiarks}, \citenamefont {Rempe},\ and\ \citenamefont
  {D\"{u}rr}}]{Baur14}%
  \BibitemOpen
  \bibfield  {author} {\bibinfo {author} {\bibfnamefont {S.}~\bibnamefont
  {Baur}}, \bibinfo {author} {\bibfnamefont {D.}~\bibnamefont {Tiarks}},
  \bibinfo {author} {\bibfnamefont {G.}~\bibnamefont {Rempe}}, \ and\ \bibinfo
  {author} {\bibfnamefont {S.}~\bibnamefont {D\"{u}rr}},\ }\bibfield  {title}
  {\enquote {\bibinfo {title} {Single-photon switch based on rydberg
  blockade},}\ }\href@noop {} {\bibfield  {journal} {\bibinfo  {journal} {Phys.
  Rev. Lett.}\ }\textbf {\bibinfo {volume} {112}},\ \bibinfo {pages} {073901}
  (\bibinfo {year} {2014})}\BibitemShut {NoStop}%
\bibitem [{\citenamefont {Tiarks}\ \emph {et~al.}(2014)\citenamefont {Tiarks},
  \citenamefont {Baur}, \citenamefont {Schneider}, \citenamefont {D\"{u}rr},\
  and\ \citenamefont {Rempe}}]{Tiarks14}%
  \BibitemOpen
  \bibfield  {author} {\bibinfo {author} {\bibfnamefont {D.}~\bibnamefont
  {Tiarks}}, \bibinfo {author} {\bibfnamefont {S.}~\bibnamefont {Baur}},
  \bibinfo {author} {\bibfnamefont {K.}~\bibnamefont {Schneider}}, \bibinfo
  {author} {\bibfnamefont {S.}~\bibnamefont {D\"{u}rr}}, \ and\ \bibinfo
  {author} {\bibfnamefont {G.}~\bibnamefont {Rempe}},\ }\bibfield  {title}
  {\enquote {\bibinfo {title} {Single-photon transistor using a f\"{o}rster
  resonance},}\ }\href@noop {} {\bibfield  {journal} {\bibinfo  {journal}
  {Phys. Rev. Lett.}\ }\textbf {\bibinfo {volume} {113}},\ \bibinfo {pages}
  {053602} (\bibinfo {year} {2014})}\BibitemShut {NoStop}%
\bibitem [{\citenamefont {Liu}\ \emph {et~al.}(2001)\citenamefont {Liu},
  \citenamefont {Dutton}, \citenamefont {Behroozi},\ and\ \citenamefont
  {Hau}}]{Liu01}%
  \BibitemOpen
  \bibfield  {author} {\bibinfo {author} {\bibfnamefont {C.}~\bibnamefont
  {Liu}}, \bibinfo {author} {\bibfnamefont {Z.}~\bibnamefont {Dutton}},
  \bibinfo {author} {\bibfnamefont {C.~H.}\ \bibnamefont {Behroozi}}, \ and\
  \bibinfo {author} {\bibfnamefont {L.~V.}\ \bibnamefont {Hau}},\ }\bibfield
  {title} {\enquote {\bibinfo {title} {Observation of coherent optical
  information storage in an atomic medium using halted light pulses},}\
  }\href@noop {} {\bibfield  {journal} {\bibinfo  {journal} {Nature}\ }\textbf
  {\bibinfo {volume} {409}},\ \bibinfo {pages} {490--493} (\bibinfo {year}
  {2001})}\BibitemShut {NoStop}%
\bibitem [{\citenamefont {Phillips}\ \emph {et~al.}(2001)\citenamefont
  {Phillips}, \citenamefont {Fleischhauer}, \citenamefont {Mair}, \citenamefont
  {Walsworth},\ and\ \citenamefont {Lukin}}]{Phillips01}%
  \BibitemOpen
  \bibfield  {author} {\bibinfo {author} {\bibfnamefont {D.~F.}\ \bibnamefont
  {Phillips}}, \bibinfo {author} {\bibfnamefont {A.}~\bibnamefont
  {Fleischhauer}}, \bibinfo {author} {\bibfnamefont {A.}~\bibnamefont {Mair}},
  \bibinfo {author} {\bibfnamefont {R.~L.}\ \bibnamefont {Walsworth}}, \ and\
  \bibinfo {author} {\bibfnamefont {M.~D.}\ \bibnamefont {Lukin}},\ }\bibfield
  {title} {\enquote {\bibinfo {title} {Storage of light in atomic vapor},}\
  }\href@noop {} {\bibfield  {journal} {\bibinfo  {journal} {Phys. Rev. Lett.}\
  }\textbf {\bibinfo {volume} {86}},\ \bibinfo {pages} {783} (\bibinfo {year}
  {2001})}\BibitemShut {NoStop}%
\bibitem [{\citenamefont {Cho}\ \emph {et~al.}(2016)\citenamefont {Cho},
  \citenamefont {Campbell}, \citenamefont {Everett}, \citenamefont {Bernu},
  \citenamefont {Higginbottom}, \citenamefont {Cao}, \citenamefont {Geng},
  \citenamefont {Robins}, \citenamefont {Lam},\ and\ \citenamefont
  {Buchler}}]{Cho16}%
  \BibitemOpen
  \bibfield  {author} {\bibinfo {author} {\bibfnamefont {Y.~W.}\ \bibnamefont
  {Cho}}, \bibinfo {author} {\bibfnamefont {G.~T.}\ \bibnamefont {Campbell}},
  \bibinfo {author} {\bibfnamefont {J.~L.}\ \bibnamefont {Everett}}, \bibinfo
  {author} {\bibfnamefont {J.}~\bibnamefont {Bernu}}, \bibinfo {author}
  {\bibfnamefont {D.~B.}\ \bibnamefont {Higginbottom}}, \bibinfo {author}
  {\bibfnamefont {M.~T.}\ \bibnamefont {Cao}}, \bibinfo {author} {\bibfnamefont
  {J.}~\bibnamefont {Geng}}, \bibinfo {author} {\bibfnamefont {N.~P.}\
  \bibnamefont {Robins}}, \bibinfo {author} {\bibfnamefont {P.~K.}\
  \bibnamefont {Lam}}, \ and\ \bibinfo {author} {\bibfnamefont {B.~C.}\
  \bibnamefont {Buchler}},\ }\bibfield  {title} {\enquote {\bibinfo {title}
  {Highly efficient optical quantum memory with long coherence time in cold
  atoms},}\ }\href@noop {} {\bibfield  {journal} {\bibinfo  {journal} {Optica}\
  }\textbf {\bibinfo {volume} {3}},\ \bibinfo {pages} {100--107} (\bibinfo
  {year} {2016})}\BibitemShut {NoStop}%
\bibitem [{\citenamefont {Fleischhauer}\ and\ \citenamefont
  {Lukin}(2002)}]{Fleishhauer02}%
  \BibitemOpen
  \bibfield  {author} {\bibinfo {author} {\bibfnamefont {M.}~\bibnamefont
  {Fleischhauer}}\ and\ \bibinfo {author} {\bibfnamefont {M.~D.}\ \bibnamefont
  {Lukin}},\ }\bibfield  {title} {\enquote {\bibinfo {title} {Quantum memory
  for photons: Dark-state polaritons},}\ }\href@noop {} {\bibfield  {journal}
  {\bibinfo  {journal} {Phys. Rev. A}\ }\textbf {\bibinfo {volume} {65}},\
  \bibinfo {pages} {022314} (\bibinfo {year} {2002})}\BibitemShut {NoStop}%
\bibitem [{\citenamefont {Wu}, \citenamefont {Artoni},\ and\ \citenamefont
  {Rocca}(2010)}]{Wu10}%
  \BibitemOpen
  \bibfield  {author} {\bibinfo {author} {\bibfnamefont {J.~H.}\ \bibnamefont
  {Wu}}, \bibinfo {author} {\bibfnamefont {M.}~\bibnamefont {Artoni}}, \ and\
  \bibinfo {author} {\bibfnamefont {G.~C.~L.}\ \bibnamefont {Rocca}},\
  }\bibfield  {title} {\enquote {\bibinfo {title} {Stationary light pulses in
  cold thermal atomic clouds},}\ }\href@noop {} {\bibfield  {journal} {\bibinfo
   {journal} {Phys. Rev. A}\ }\textbf {\bibinfo {volume} {82}},\ \bibinfo
  {pages} {013807} (\bibinfo {year} {2010})}\BibitemShut {NoStop}%
\bibitem [{\citenamefont {Andr\'{e}}\ \emph {et~al.}(2005)\citenamefont
  {Andr\'{e}}, \citenamefont {Bajcsy}, \citenamefont {Zibrov},\ and\
  \citenamefont {Lukin}}]{Andre09}%
  \BibitemOpen
  \bibfield  {author} {\bibinfo {author} {\bibfnamefont {A.}~\bibnamefont
  {Andr\'{e}}}, \bibinfo {author} {\bibfnamefont {M.}~\bibnamefont {Bajcsy}},
  \bibinfo {author} {\bibfnamefont {A.~S.}\ \bibnamefont {Zibrov}}, \ and\
  \bibinfo {author} {\bibfnamefont {M.~D.}\ \bibnamefont {Lukin}},\ }\bibfield
  {title} {\enquote {\bibinfo {title} {Nonlinear optics with stationary pulses
  of light},}\ }\href@noop {} {\bibfield  {journal} {\bibinfo  {journal} {Phys.
  Rev. Lett.}\ }\textbf {\bibinfo {volume} {94}},\ \bibinfo {pages} {063902}
  (\bibinfo {year} {2005})}\BibitemShut {NoStop}%
\bibitem [{\citenamefont {Lin}\ \emph {et~al.}(2009)\citenamefont {Lin},
  \citenamefont {Liao}, \citenamefont {Peters}, \citenamefont {Chou},
  \citenamefont {Wang}, \citenamefont {Cho}, \citenamefont {Kuan},\ and\
  \citenamefont {Yu}}]{Lin09}%
  \BibitemOpen
  \bibfield  {author} {\bibinfo {author} {\bibfnamefont {Y.~W.}\ \bibnamefont
  {Lin}}, \bibinfo {author} {\bibfnamefont {W.~T.}\ \bibnamefont {Liao}},
  \bibinfo {author} {\bibfnamefont {T.}~\bibnamefont {Peters}}, \bibinfo
  {author} {\bibfnamefont {H.~C.}\ \bibnamefont {Chou}}, \bibinfo {author}
  {\bibfnamefont {J.~S.}\ \bibnamefont {Wang}}, \bibinfo {author}
  {\bibfnamefont {H.~W.}\ \bibnamefont {Cho}}, \bibinfo {author} {\bibfnamefont
  {P.~C.}\ \bibnamefont {Kuan}}, \ and\ \bibinfo {author} {\bibfnamefont
  {I.~A.}\ \bibnamefont {Yu}},\ }\bibfield  {title} {\enquote {\bibinfo {title}
  {Stationary light pulses in cold atomic media and without bragg gratings},}\
  }\href@noop {} {\bibfield  {journal} {\bibinfo  {journal} {Phys. Rev. Lett.}\
  }\textbf {\bibinfo {volume} {102}},\ \bibinfo {pages} {213601} (\bibinfo
  {year} {2009})}\BibitemShut {NoStop}%
\bibitem [{\citenamefont {Everett}\ \emph {et~al.}(2017)\citenamefont
  {Everett}, \citenamefont {Campbell}, \citenamefont {Cho}, \citenamefont
  {Vernaz-Gris}, \citenamefont {Higginbottom}, \citenamefont {Pinel},
  \citenamefont {Robins}, \citenamefont {Lam},\ and\ \citenamefont
  {Buchler}}]{Everett17}%
  \BibitemOpen
  \bibfield  {author} {\bibinfo {author} {\bibfnamefont {J.~L.}\ \bibnamefont
  {Everett}}, \bibinfo {author} {\bibfnamefont {G.~T.}\ \bibnamefont
  {Campbell}}, \bibinfo {author} {\bibfnamefont {Y.~W.}\ \bibnamefont {Cho}},
  \bibinfo {author} {\bibfnamefont {P.}~\bibnamefont {Vernaz-Gris}}, \bibinfo
  {author} {\bibfnamefont {D.~B.}\ \bibnamefont {Higginbottom}}, \bibinfo
  {author} {\bibfnamefont {O.}~\bibnamefont {Pinel}}, \bibinfo {author}
  {\bibfnamefont {N.~P.}\ \bibnamefont {Robins}}, \bibinfo {author}
  {\bibfnamefont {P.~K.}\ \bibnamefont {Lam}}, \ and\ \bibinfo {author}
  {\bibfnamefont {B.~C.}\ \bibnamefont {Buchler}},\ }\bibfield  {title}
  {\enquote {\bibinfo {title} {Dynamical observation of self-stabilizing
  stationary light},}\ }\href@noop {} {\bibfield  {journal} {\bibinfo
  {journal} {Nat. Phys.}\ }\textbf {\bibinfo {volume} {13}},\ \bibinfo {pages}
  {68--73} (\bibinfo {year} {2017})}\BibitemShut {NoStop}%
\bibitem [{\citenamefont {Bajcsy}, \citenamefont {Zibrov},\ and\ \citenamefont
  {Lukin}(2003)}]{Bajcsy03}%
  \BibitemOpen
  \bibfield  {author} {\bibinfo {author} {\bibfnamefont {M.}~\bibnamefont
  {Bajcsy}}, \bibinfo {author} {\bibfnamefont {A.~S.}\ \bibnamefont {Zibrov}},
  \ and\ \bibinfo {author} {\bibfnamefont {M.~D.}\ \bibnamefont {Lukin}},\
  }\bibfield  {title} {\enquote {\bibinfo {title} {Stationary pulses of light
  in an atomic medium},}\ }\href@noop {} {\bibfield  {journal} {\bibinfo
  {journal} {Nature}\ }\textbf {\bibinfo {volume} {426}},\ \bibinfo {pages}
  {638--641} (\bibinfo {year} {2003})}\BibitemShut {NoStop}%
\bibitem [{\citenamefont {Chen}\ \emph {et~al.}(2012)\citenamefont {Chen},
  \citenamefont {Lee}, \citenamefont {Hung}, \citenamefont {Chen},
  \citenamefont {Chen},\ and\ \citenamefont {Yu}}]{Chen12}%
  \BibitemOpen
  \bibfield  {author} {\bibinfo {author} {\bibfnamefont {Y.-H.}\ \bibnamefont
  {Chen}}, \bibinfo {author} {\bibfnamefont {M.-J.}\ \bibnamefont {Lee}},
  \bibinfo {author} {\bibfnamefont {W.}~\bibnamefont {Hung}}, \bibinfo {author}
  {\bibfnamefont {Y.-C.}\ \bibnamefont {Chen}}, \bibinfo {author}
  {\bibfnamefont {Y.-F.}\ \bibnamefont {Chen}}, \ and\ \bibinfo {author}
  {\bibfnamefont {I.~A.}\ \bibnamefont {Yu}},\ }\bibfield  {title} {\enquote
  {\bibinfo {title} {Demonstration of the interaction between two stopped light
  pulses},}\ }\href@noop {} {\bibfield  {journal} {\bibinfo  {journal} {Phys.
  Rev. Lett.}\ }\textbf {\bibinfo {volume} {108}},\ \bibinfo {pages} {173603}
  (\bibinfo {year} {2012})}\BibitemShut {NoStop}%
\bibitem [{\citenamefont {Park}\ \emph {et~al.}(2018)\citenamefont {Park},
  \citenamefont {Cho}, \citenamefont {Chough},\ and\ \citenamefont
  {Kim}}]{Park18}%
  \BibitemOpen
  \bibfield  {author} {\bibinfo {author} {\bibfnamefont {K.-K.}\ \bibnamefont
  {Park}}, \bibinfo {author} {\bibfnamefont {Y.-W.}\ \bibnamefont {Cho}},
  \bibinfo {author} {\bibfnamefont {Y.-T.}\ \bibnamefont {Chough}}, \ and\
  \bibinfo {author} {\bibfnamefont {Y.-H.}\ \bibnamefont {Kim}},\ }\bibfield
  {title} {\enquote {\bibinfo {title} {Experimental demonstration of quantum
  stationary light pulses in an atomic ensemble},}\ }\href@noop {} {\bibfield
  {journal} {\bibinfo  {journal} {Phys. Rev. X}\ }\textbf {\bibinfo {volume}
  {8}},\ \bibinfo {pages} {021016} (\bibinfo {year} {2018})}\BibitemShut
  {NoStop}%
\bibitem [{\citenamefont {Iakoupov}, \citenamefont {Borregaard},\ and\
  \citenamefont {rensen}(2018)}]{Iakoupov18}%
  \BibitemOpen
  \bibfield  {author} {\bibinfo {author} {\bibfnamefont {I.}~\bibnamefont
  {Iakoupov}}, \bibinfo {author} {\bibfnamefont {J.}~\bibnamefont
  {Borregaard}}, \ and\ \bibinfo {author} {\bibfnamefont {A.~S.~S.}\
  \bibnamefont {rensen}},\ }\bibfield  {title} {\enquote {\bibinfo {title}
  {Controlled-phase gate for photons based on stationary light},}\ }\href@noop
  {} {\bibfield  {journal} {\bibinfo  {journal} {Phys. Rev. Lett.}\ }\textbf
  {\bibinfo {volume} {120}},\ \bibinfo {pages} {010502} (\bibinfo {year}
  {2018})}\BibitemShut {NoStop}%
\bibitem [{\citenamefont {Chang}\ \emph {et~al.}(2008)\citenamefont {Chang},
  \citenamefont {Gritsev}, \citenamefont {Morigi}, \citenamefont {Vuleti\'{c}},
  \citenamefont {Lukin},\ and\ \citenamefont {Demler}}]{Chang08}%
  \BibitemOpen
  \bibfield  {author} {\bibinfo {author} {\bibfnamefont {D.~E.}\ \bibnamefont
  {Chang}}, \bibinfo {author} {\bibfnamefont {V.}~\bibnamefont {Gritsev}},
  \bibinfo {author} {\bibfnamefont {G.}~\bibnamefont {Morigi}}, \bibinfo
  {author} {\bibfnamefont {V.}~\bibnamefont {Vuleti\'{c}}}, \bibinfo {author}
  {\bibfnamefont {M.~D.}\ \bibnamefont {Lukin}}, \ and\ \bibinfo {author}
  {\bibfnamefont {E.~A.}\ \bibnamefont {Demler}},\ }\bibfield  {title}
  {\enquote {\bibinfo {title} {rystallization of strongly interacting photons
  in a nonlinear optical fibre},}\ }\href@noop {} {\bibfield  {journal}
  {\bibinfo  {journal} {Nat. Phys.}\ }\textbf {\bibinfo {volume} {4}},\
  \bibinfo {pages} {884--889} (\bibinfo {year} {2008})}\BibitemShut {NoStop}%
\bibitem [{\citenamefont {Fleischhauer}, \citenamefont {Otterbach},\ and\
  \citenamefont {Unanyan}(2008)}]{Fleischhauer08}%
  \BibitemOpen
  \bibfield  {author} {\bibinfo {author} {\bibfnamefont {M.}~\bibnamefont
  {Fleischhauer}}, \bibinfo {author} {\bibfnamefont {J.}~\bibnamefont
  {Otterbach}}, \ and\ \bibinfo {author} {\bibfnamefont {R.~G.}\ \bibnamefont
  {Unanyan}},\ }\bibfield  {title} {\enquote {\bibinfo {title} {Bose-einstein
  condensation of stationary-light polaritons},}\ }\href@noop {} {\bibfield
  {journal} {\bibinfo  {journal} {Phys. Rev. Lett.}\ }\textbf {\bibinfo
  {volume} {101}},\ \bibinfo {pages} {163601} (\bibinfo {year}
  {2008})}\BibitemShut {NoStop}%
\bibitem [{\citenamefont {Angelakis}\ \emph {et~al.}(2013)\citenamefont
  {Angelakis}, \citenamefont {Huo}, \citenamefont {Chang}, \citenamefont
  {Kwek},\ and\ \citenamefont {Korepin}}]{Dimitris13}%
  \BibitemOpen
  \bibfield  {author} {\bibinfo {author} {\bibfnamefont {D.~G.}\ \bibnamefont
  {Angelakis}}, \bibinfo {author} {\bibfnamefont {M.-X.}\ \bibnamefont {Huo}},
  \bibinfo {author} {\bibfnamefont {D.}~\bibnamefont {Chang}}, \bibinfo
  {author} {\bibfnamefont {L.~C.}\ \bibnamefont {Kwek}}, \ and\ \bibinfo
  {author} {\bibfnamefont {V.}~\bibnamefont {Korepin}},\ }\bibfield  {title}
  {\enquote {\bibinfo {title} {Mimicking interacting relativistic theories with
  stationary pulses of light},}\ }\href@noop {} {\bibfield  {journal} {\bibinfo
   {journal} {Phys. Rev. Lett.}\ }\textbf {\bibinfo {volume} {110}},\ \bibinfo
  {pages} {100502} (\bibinfo {year} {2013})}\BibitemShut {NoStop}%
\bibitem [{\citenamefont {Cho}\ \emph {et~al.}(2014)\citenamefont {Cho},
  \citenamefont {Park}, \citenamefont {Lee},\ and\ \citenamefont
  {Kim}}]{Cho14}%
  \BibitemOpen
  \bibfield  {author} {\bibinfo {author} {\bibfnamefont {Y.-W.}\ \bibnamefont
  {Cho}}, \bibinfo {author} {\bibfnamefont {K.-K.}\ \bibnamefont {Park}},
  \bibinfo {author} {\bibfnamefont {J.-C.}\ \bibnamefont {Lee}}, \ and\
  \bibinfo {author} {\bibfnamefont {Y.-H.}\ \bibnamefont {Kim}},\ }\bibfield
  {title} {\enquote {\bibinfo {title} {Engineering frequency-time quantum
  correlation of narrow-band biphotons from cold atoms},}\ }\href@noop {}
  {\bibfield  {journal} {\bibinfo  {journal} {Phys. Rev. Lett.}\ }\textbf
  {\bibinfo {volume} {113}},\ \bibinfo {pages} {063602} (\bibinfo {year}
  {2014})}\BibitemShut {NoStop}%
\bibitem [{\citenamefont {Lee}\ \emph {et~al.}(2016)\citenamefont {Lee},
  \citenamefont {Park}, \citenamefont {Zhao},\ and\ \citenamefont
  {Kim}}]{Lee16}%
  \BibitemOpen
  \bibfield  {author} {\bibinfo {author} {\bibfnamefont {J.-C.}\ \bibnamefont
  {Lee}}, \bibinfo {author} {\bibfnamefont {K.-K.}\ \bibnamefont {Park}},
  \bibinfo {author} {\bibfnamefont {T.-M.}\ \bibnamefont {Zhao}}, \ and\
  \bibinfo {author} {\bibfnamefont {Y.-H.}\ \bibnamefont {Kim}},\ }\bibfield
  {title} {\enquote {\bibinfo {title} {Einstein-podolsky-rosen entanglement of
  narrow-band photons from cold atoms},}\ }\href@noop {} {\bibfield  {journal}
  {\bibinfo  {journal} {Phys. Rev. Lett.}\ }\textbf {\bibinfo {volume} {117}},\
  \bibinfo {pages} {250501} (\bibinfo {year} {2016})}\BibitemShut {NoStop}%
\bibitem [{\citenamefont {Zhao}, \citenamefont {Ihn},\ and\ \citenamefont
  {Kim}(2019)}]{Ming19}%
  \BibitemOpen
  \bibfield  {author} {\bibinfo {author} {\bibfnamefont {T.-M.}\ \bibnamefont
  {Zhao}}, \bibinfo {author} {\bibfnamefont {Y.~S.}\ \bibnamefont {Ihn}}, \
  and\ \bibinfo {author} {\bibfnamefont {Y.-H.}\ \bibnamefont {Kim}},\
  }\bibfield  {title} {\enquote {\bibinfo {title} {Direct generation of
  narrow-band hyperentangled photons},}\ }\href@noop {} {\bibfield  {journal}
  {\bibinfo  {journal} {Phys. Rev. Lett.}\ }\textbf {\bibinfo {volume} {122}},\
  \bibinfo {pages} {123607} (\bibinfo {year} {2019})}\BibitemShut {NoStop}%
\bibitem [{\citenamefont {Park}\ \emph {et~al.}(2017)\citenamefont {Park},
  \citenamefont {Kim}, \citenamefont {Zhao}, \citenamefont {Cho},\ and\
  \citenamefont {Kim}}]{Park17}%
  \BibitemOpen
  \bibfield  {author} {\bibinfo {author} {\bibfnamefont {K.-K.}\ \bibnamefont
  {Park}}, \bibinfo {author} {\bibfnamefont {J.-H.}\ \bibnamefont {Kim}},
  \bibinfo {author} {\bibfnamefont {T.-M.}\ \bibnamefont {Zhao}}, \bibinfo
  {author} {\bibfnamefont {Y.-W.}\ \bibnamefont {Cho}}, \ and\ \bibinfo
  {author} {\bibfnamefont {Y.-H.}\ \bibnamefont {Kim}},\ }\bibfield  {title}
  {\enquote {\bibinfo {title} {Measuring the frequency-time two-photon
  wavefunction of narrowband entangled photons from cold atoms via stimulated
  emission},}\ }\href@noop {} {\bibfield  {journal} {\bibinfo  {journal}
  {Optica}\ }\textbf {\bibinfo {volume} {4}},\ \bibinfo {pages} {1293--1297}
  (\bibinfo {year} {2017})}\BibitemShut {NoStop}%
\bibitem [{\citenamefont {Zhao}, \citenamefont {Su},\ and\ \citenamefont
  {Du}(2016)}]{du16}%
  \BibitemOpen
  \bibfield  {author} {\bibinfo {author} {\bibfnamefont {L.}~\bibnamefont
  {Zhao}}, \bibinfo {author} {\bibfnamefont {Y.}~\bibnamefont {Su}}, \ and\
  \bibinfo {author} {\bibfnamefont {S.}~\bibnamefont {Du}},\ }\bibfield
  {title} {\enquote {\bibinfo {title} {Narrowband biphoton generation in the
  group delay regime},}\ }\href@noop {} {\bibfield  {journal} {\bibinfo
  {journal} {Phys. Rev. A}\ }\textbf {\bibinfo {volume} {93}},\ \bibinfo
  {pages} {033815} (\bibinfo {year} {2016})}\BibitemShut {NoStop}%
\bibitem [{\citenamefont {Moiseev}\ and\ \citenamefont
  {Ham}(2006)}]{Moiseev06}%
  \BibitemOpen
  \bibfield  {author} {\bibinfo {author} {\bibfnamefont {S.~A.}\ \bibnamefont
  {Moiseev}}\ and\ \bibinfo {author} {\bibfnamefont {B.~S.}\ \bibnamefont
  {Ham}},\ }\bibfield  {title} {\enquote {\bibinfo {title} {Quantum
  manipulation of two-color stationary light: Quantum wavelength conversion},}\
  }\href@noop {} {\bibfield  {journal} {\bibinfo  {journal} {Phys. Rev. A}\
  }\textbf {\bibinfo {volume} {73}},\ \bibinfo {pages} {033812} (\bibinfo
  {year} {2006})}\BibitemShut {NoStop}%
\bibitem [{\citenamefont {Iakoupov}, \citenamefont {J.~R.~Ott},\ and\
  \citenamefont {rensen}(2016)}]{Iakoupov16}%
  \BibitemOpen
  \bibfield  {author} {\bibinfo {author} {\bibfnamefont {I.}~\bibnamefont
  {Iakoupov}}, \bibinfo {author} {\bibfnamefont {D.~E.~C.}\ \bibnamefont
  {J.~R.~Ott}}, \ and\ \bibinfo {author} {\bibfnamefont {A.~S.~S.}\
  \bibnamefont {rensen}},\ }\bibfield  {title} {\enquote {\bibinfo {title}
  {Dispersion relations for stationary light in one-dimensional atomic
  ensembles},}\ }\href@noop {} {\bibfield  {journal} {\bibinfo  {journal}
  {Phys. Rev. A}\ }\textbf {\bibinfo {volume} {94}},\ \bibinfo {pages} {053824}
  (\bibinfo {year} {2016})}\BibitemShut {NoStop}%
\bibitem [{\citenamefont {Zimmer}\ \emph {et~al.}(2008)\citenamefont {Zimmer},
  \citenamefont {Otterbach}, \citenamefont {Unanyan}, \citenamefont {Shore},\
  and\ \citenamefont {Fleischhauer}}]{Zimmer08}%
  \BibitemOpen
  \bibfield  {author} {\bibinfo {author} {\bibfnamefont {F.~E.}\ \bibnamefont
  {Zimmer}}, \bibinfo {author} {\bibfnamefont {J.}~\bibnamefont {Otterbach}},
  \bibinfo {author} {\bibfnamefont {R.~G.}\ \bibnamefont {Unanyan}}, \bibinfo
  {author} {\bibfnamefont {B.~W.}\ \bibnamefont {Shore}}, \ and\ \bibinfo
  {author} {\bibfnamefont {M.}~\bibnamefont {Fleischhauer}},\ }\bibfield
  {title} {\enquote {\bibinfo {title} {Dark-state polaritons for multicomponent
  and stationary light fields},}\ }\href@noop {} {\bibfield  {journal}
  {\bibinfo  {journal} {Phys. Rev. A}\ }\textbf {\bibinfo {volume} {77}},\
  \bibinfo {pages} {063823} (\bibinfo {year} {2008})}\BibitemShut {NoStop}%
\bibitem [{\citenamefont {Fleischhauer}\ and\ \citenamefont
  {Lukin}(2000)}]{Fleischhauer00}%
  \BibitemOpen
  \bibfield  {author} {\bibinfo {author} {\bibfnamefont {M.}~\bibnamefont
  {Fleischhauer}}\ and\ \bibinfo {author} {\bibfnamefont {M.~D.}\ \bibnamefont
  {Lukin}},\ }\bibfield  {title} {\enquote {\bibinfo {title} {Dark-state
  polaritons in electromagnetically induced transparency},}\ }\href@noop {}
  {\bibfield  {journal} {\bibinfo  {journal} {Phys. Rev. Lett.}\ }\textbf
  {\bibinfo {volume} {84}},\ \bibinfo {pages} {5094} (\bibinfo {year}
  {2000})}\BibitemShut {NoStop}%
\bibitem [{\citenamefont {Hansen}\ and\ \citenamefont {lmer}(2007)}]{Hansen07}%
  \BibitemOpen
  \bibfield  {author} {\bibinfo {author} {\bibfnamefont {K.~R.}\ \bibnamefont
  {Hansen}}\ and\ \bibinfo {author} {\bibfnamefont {K.~M.}\ \bibnamefont
  {lmer}},\ }\bibfield  {title} {\enquote {\bibinfo {title} {Trapping of light
  pulses in ensembles of stationary $\lambda$ atoms},}\ }\href@noop {}
  {\bibfield  {journal} {\bibinfo  {journal} {Phys. Rev. A}\ }\textbf {\bibinfo
  {volume} {75}},\ \bibinfo {pages} {053802} (\bibinfo {year}
  {2007})}\BibitemShut {NoStop}%
\bibitem [{\citenamefont {Wang}\ \emph {et~al.}(2019)\citenamefont {Wang},
  \citenamefont {Li}, \citenamefont {Zhang}, \citenamefont {Su}, \citenamefont
  {Zhou}, \citenamefont {Liao}, \citenamefont {Du}, \citenamefont {Yan},\ and\
  \citenamefont {Zhu}}]{Wang19}%
  \BibitemOpen
  \bibfield  {author} {\bibinfo {author} {\bibfnamefont {Y.}~\bibnamefont
  {Wang}}, \bibinfo {author} {\bibfnamefont {J.}~\bibnamefont {Li}}, \bibinfo
  {author} {\bibfnamefont {S.}~\bibnamefont {Zhang}}, \bibinfo {author}
  {\bibfnamefont {K.}~\bibnamefont {Su}}, \bibinfo {author} {\bibfnamefont
  {Y.}~\bibnamefont {Zhou}}, \bibinfo {author} {\bibfnamefont {K.}~\bibnamefont
  {Liao}}, \bibinfo {author} {\bibfnamefont {S.}~\bibnamefont {Du}}, \bibinfo
  {author} {\bibfnamefont {H.}~\bibnamefont {Yan}}, \ and\ \bibinfo {author}
  {\bibfnamefont {S.~L.}\ \bibnamefont {Zhu}},\ }\bibfield  {title} {\enquote
  {\bibinfo {title} {Efficient quantum memory for single-photon polarization
  qubits},}\ }\href@noop {} {\bibfield  {journal} {\bibinfo  {journal} {Nat.
  Photonics}\ }\textbf {\bibinfo {volume} {13}},\ \bibinfo {pages} {346--351}
  (\bibinfo {year} {2019})}\BibitemShut {NoStop}%
\bibitem [{\citenamefont {Fan}\ and\ \citenamefont {Yu}(2002)}]{Hongyi02}%
  \BibitemOpen
  \bibfield  {author} {\bibinfo {author} {\bibfnamefont {H.}~\bibnamefont
  {Fan}}\ and\ \bibinfo {author} {\bibfnamefont {G.}~\bibnamefont {Yu}},\
  }\bibfield  {title} {\enquote {\bibinfo {title} {Three-mode squeezed vacuum
  state in fock space as an entangled state},}\ }\href@noop {} {\bibfield
  {journal} {\bibinfo  {journal} {Phys. Rev. A}\ }\textbf {\bibinfo {volume}
  {65}},\ \bibinfo {pages} {033829} (\bibinfo {year} {2002})}\BibitemShut
  {NoStop}%
\bibitem [{\citenamefont {Cochrane}, \citenamefont {Ralph},\ and\ \citenamefont
  {Milburn}(2002)}]{Cochrane02}%
  \BibitemOpen
  \bibfield  {author} {\bibinfo {author} {\bibfnamefont {P.~T.}\ \bibnamefont
  {Cochrane}}, \bibinfo {author} {\bibfnamefont {T.~C.}\ \bibnamefont {Ralph}},
  \ and\ \bibinfo {author} {\bibfnamefont {G.~J.}\ \bibnamefont {Milburn}},\
  }\bibfield  {title} {\enquote {\bibinfo {title} {Teleportation improvement by
  conditional measurements on the two-mode squeezed vacuum},}\ }\href@noop {}
  {\bibfield  {journal} {\bibinfo  {journal} {Phys. Rev. A}\ }\textbf {\bibinfo
  {volume} {65}},\ \bibinfo {pages} {062306} (\bibinfo {year}
  {2002})}\BibitemShut {NoStop}%
\bibitem [{\citenamefont {Braje}\ \emph {et~al.}(2004)\citenamefont {Braje},
  \citenamefont {Bali\'{c}}, \citenamefont {Goda}, \citenamefont {Yin},\ and\
  \citenamefont {Harris}}]{Danielle04}%
  \BibitemOpen
  \bibfield  {author} {\bibinfo {author} {\bibfnamefont {D.~A.}\ \bibnamefont
  {Braje}}, \bibinfo {author} {\bibfnamefont {V.}~\bibnamefont {Bali\'{c}}},
  \bibinfo {author} {\bibfnamefont {S.}~\bibnamefont {Goda}}, \bibinfo {author}
  {\bibfnamefont {G.~Y.}\ \bibnamefont {Yin}}, \ and\ \bibinfo {author}
  {\bibfnamefont {S.~E.}\ \bibnamefont {Harris}},\ }\bibfield  {title}
  {\enquote {\bibinfo {title} {Frequency mixing using electromagnetically
  induced transparency in cold atoms},}\ }\href@noop {} {\bibfield  {journal}
  {\bibinfo  {journal} {Phys. Rev. Lett.}\ }\textbf {\bibinfo {volume} {93}},\
  \bibinfo {pages} {183601} (\bibinfo {year} {2004})}\BibitemShut {NoStop}%
\bibitem [{\citenamefont {Dehghan}(1999)}]{Dehghan99}%
  \BibitemOpen
  \bibfield  {author} {\bibinfo {author} {\bibfnamefont {M.}~\bibnamefont
  {Dehghan}},\ }\bibfield  {title} {\enquote {\bibinfo {title} {Fully implicit
  finite differences methods for two-dimensional diffusion with a non-local
  boundary condition},}\ }\href@noop {} {\bibfield  {journal} {\bibinfo
  {journal} {J. Comput. Appl. Math.}\ }\textbf {\bibinfo {volume} {106}},\
  \bibinfo {pages} {255--269} (\bibinfo {year} {1999})}\BibitemShut {NoStop}%
\bibitem [{\citenamefont {Curtiss}\ and\ \citenamefont
  {Hirschfelder}(1952)}]{Curtiss52}%
  \BibitemOpen
  \bibfield  {author} {\bibinfo {author} {\bibfnamefont {C.~F.}\ \bibnamefont
  {Curtiss}}\ and\ \bibinfo {author} {\bibfnamefont {J.~O.}\ \bibnamefont
  {Hirschfelder}},\ }\bibfield  {title} {\enquote {\bibinfo {title}
  {Integration of stiff equation},}\ }\href@noop {} {\bibfield  {journal}
  {\bibinfo  {journal} {Proc. Natl. Acad. Sci. U.S.A.}\ }\textbf {\bibinfo
  {volume} {38}},\ \bibinfo {pages} {235--243} (\bibinfo {year}
  {1952})}\BibitemShut {NoStop}%
\end{thebibliography}%

\end{document}